\documentclass{aastex}
\usepackage{graphicx}

\newcommand{\etal}{{\it et al}.\ }

\slugcomment{Submitted to {\it Icarus}}

\shorttitle{Physical and orbital properties of the Trojan asteroids}
\shortauthors{Melita {\it et al.\ }}

\begin{document}

\title{Physical properties and orbital stability of the Trojan asteroids} 

\author{ Melita, M.D.$^1$ \and Licandro, J.$^2$ \and Jones, D.C.$^3$ \and
Williams, I.P.$^3$ \\ $^1$ IAFE (UBA, CONICET) Argentina.
melita@iafe.uba.ar\\ $^2$ Isaac Newton Group of Telescopes, La Palma, Spain \and
Instituto de Astrof\'isica de Canarias, Tenerife, Spain.\\ $^3$
Astronomy Unit, Queen Mary College, University of London, UK.\\}

\date{}

\begin{abstract}

All the Trojan asteroids orbit about the Sun at roughly the same
heliocentric distance as Jupiter.  Differences in the observed visible
reflection spectra range from neutral to red, with no ultra-red objects
found so far. Given that the Trojan asteroids are collisionally evolved, a
certain degree of variability is expected. Additionally, cosmic radiation
and sublimation are important factors in modifying icy surfaces even at
those large heliocentric distances.  

We search for correlations between physical and dynamical properties, we
explore relationships between the following four quantities; the normalised
visible reflectivity indexes ($S'$), the absolute magnitudes, the observed
albedos and the orbital stability of the Trojans. 

We present here visible spectroscopic spectra of 25 Trojans. This new data
increase by a factor of about 5 the size of the sample of visible spectra of
Jupiter Trojans on unstable orbits. The observations were carried out at the
ESO-NTT telescope (3.5m) at La Silla, Chile, the ING-WHT (4.2m) and NOT
(2.5m) at Roque de los Muchachos observatory, La Palma, Spain. 

We have found a correlation between the size distribution and the orbital
stability. The absolute-magnitude distribution of the Trojans in stable
orbits is found to be bimodal, while the one of the unstable orbits is
unimodal, with a slope similar to that of the small stable Trojans. This
supports the hypothesis that the unstable objects are mainly byproducts of
physical collisions. 

The values of $S'$ of both the stable and the unstable Trojans are uniformly
distributed over a wide range, from $0\ \%/1000\AA $ to about $15\
\%/1000\AA$. The values for the stable Trojans tend to be slightly redder
than the unstable ones, but no significant statistical difference is found.


\end{abstract}


\maketitle

\section{Introduction}


The main motivation of this work is to look for correlations between
physical and dynamical properties.%
We shall discuss ways in which the diversity in surface
properties can develop, how relationships with orbital stability may
originate and what this may imply.

The Trojan asteroids are believed to have been formed in the outer Solar
System (see for example Marzari \& Scholl 1998 or Morbidelli \etal 2005) so,
ices may be expected to be their main component. However, at the moment, no
water or other more volatile substances have been
detected on the surface of a Trojan (Yang \& Jewitt 2007, Emery \& Brown
2003, Emery \& Brown 2004). Infrared observations indicate the presence of
silicates (Cruikshank \etal 2001, Emery \etal 2006). In the case of Ennomos,
an unusually hight-albedo object, the surface-content of water ice 
has been
quantified to be below $10$\% in mass (Yang \& Jewitt 2007). 
 
Cosmic Radiation modifies the spectroscopic properties of these types of
compounds. The precise effect depends on the chemical composition of the
asteroidal surface, that of the incoming cosmic radiation and its energy.
Laboratory experiments of ion-radiation, of energies in the order of keV,
gradually `flattens' the spectral slopes of organic-complexes such as
asphaltite and kerite (Moroz et al. 2004), while it produces red and dark
residuals upon ice-surfaces such as methanol, methane and benzene, turning the
spectra to neutral for very high doses (Brunetto et al. 2006a).
Micrometeorite bombardment also lowers the albedo and reddens the surface of
silicates rich in olivine, pyroxene and serpentine, as indicated by
experiments using UV-laser pulses which simulate this effect (Brunetto et
al. 2006b); the corresponding timescale to modify the spectral slope on main
belt asteroids are estimated to range between 10$^8$yr to 10$^{10}$yr. On
the other hand, experiments of ion-irradiation on a sample of the meteorite
Epinal indicate that this timescale for an S-type asteroid could be as short
as 10$^4$yr to 10$^6$yr in the NEA population (Strazzulla et al. 2005). The
timescale in which the surface spectral properties of asteroids will be
modified in the Jupiter region is not simple to estimate as it depends on
the modulation of the solar activity at these distances. But, since the
density of the cosmic radiation originated in the Sun decreases as the
heliocentric-distance squared, on may argue that, if any of these processes
is relevant for the Trojans, these timescales should be increased by at
least an order of magnitude.

The abundance of irradiated material decreases with depth,
hence, if material originally from the interior is exposed in some objects,
color differences would emerge between members of the same population.
Collisions between objects is an obvious way of achieving this and,
according to Dell'Oro \etal (1998), collisions can be significant in the
Trojan population. In the trans-Neptunian population, Gil-Hutton (2002) has
shown that links between surface and orbital properties can be established
in this way.

Although the Trojan asteroids are located at a large heliocentric distance,
ice sublimation could also be an important factor affecting their surfaces.
A thin insulator mantle can form in a very short timescale that ranges from
$10^4$ yr to $10^6$ yr (Jewitt 2002). 
%

There is a slow dynamical evaporation of bodies from the Trojan clouds due
to gravitational perturbations, with the $5:2$ mean motion resonance with
Saturn being very important (Levison \etal 1997, Nesvorny \& Dones 2002,
Tsiganis \etal 2005). In particular, Thersites (1868), has a high
probability of leaving the Trojan swarm in less than $50 Myr$ (Tsiganis
\etal 2000). But the out-flux rate produced by collisions is much larger.
Marzari \etal (1995) have found that there is an out-flux of one Trojan
asteroid greater than $1 km$ in size every $1000 yr$. The smaller byproducts
of a collision are more numerous and tend to have greater post-encounter
velocities and are thus less likely to remain in the most stable regions.
Hence, collisions could create variability in surface properties and at the
same time produce a correlation between physical properties and dynamical
state. 

Within the Trojan Clouds there exist a number of dynamical families, which
are believed to have been formed collisionally. Fornasier \etal (2004),
Fornasier \etal (2007)  and
Dotto \etal (2006) found that, for a given family, there is a similarity in
the surface properties of the members, except for a few objects that are
physically different and these have been called {\it interlopers}.
Unfortunately, the information regarding the surfaces of Trojan asteroids
that belong to families is rather sparse to date. 



It must also be noticed that interloper objects such as captured comets
might exist in the Trojan clouds (Rabe 1972, Yoder 1979). Indeed, objects in
Centaur-type orbits are known to evolve into temporary Trojans for
time-spans of some 0.5 Myr (Horner \etal 2005) and `transitional' objects
with orbits that can evolve into short period comets are known to exist in
the $1:1$ resonance with Jupiter (Karlsonn 2004). 


Up to now, it was not possible to investigate correlations between dynamical
stability and spectroscopic properties because the colors of the
transitional Trojans were largely unknown. In order to rectify this, we have
obtained new low-resolution visible spectra of an additional $24$ Trojans in
unstable orbits.

This article is organised as follows. In section~\ref{sec:os} we discuss the
methods used to study the dynamical properties of the Trojan asteroids. %
We present the data used in this study
in section~\ref{sec:vs}, both the observations carried out for the purpose
of this investigation and also the data taken from other authors.
Correlations between physical and dynamical properties are
discussed in section~\ref{sec:corr}. 
Finally, in section~\ref{sec:disc}, we discuss our results

\section{ 
Determination of dynamical properties}
\label{sec:os}

In order to discuss whether a Trojan is in a primordial orbit or subject to
a recent perturbation we need to know the time that it can remain on or near
its present orbit. Variations in the proper elements of the orbits can occur
as a result of physical collisions. As a consequence of this random walk in
phase space, the asteroid may find an unstable regime, leading to the
expulsion from the Trojan Cloud in a timescale that is smaller than the age
of the Solar System --i.e. its present life as a Trojan is only a
`transitional' phase. If some asteroids exist in those transitional orbits
today, we may conclude that they are likely to have been inserted there
recently, most probably because of collisions (Marzari \etal 1995). 

The best way of determining the residence lifetime is to numerically
integrate the equations of motion of each asteroid over the age of the Solar
System. However, even with efficient integrators, this is a very large task.
For this reason, we have explored the accuracy of a chaos estimator, the
Lyapunov Characteristic Exponent (LCE) when applied to estimating macroscopic
stability of the orbits of known Trojan asteroids. The $LCE$ gives the rate of
exponential divergence from perturbed initial conditions, given $X_0(t)$, a
point in the orbit of the unperturbed problem, if we can write
$X(t)=X_0(t)+U(t)$, where $X(t)$ is the perturbed solution and
$U(t)$ is the deviation from the unperturbed trajectory at time
$t$, the $LCE$ is defined as:
$$LCE =lim_{t \rightarrow \infty}\ \frac{1}{t}\ ln|U(t)|.$$
Naturally, we do not expect the $LCE$ it to be $100\%$ reliable as an
estimator of macroscopic stability, because some
orbits can be in a `stable-chaos' regime (Milani and Nobili 1992, Milani
1993, Milani \etal 1997), which may render both a high value of $LCE$ and a
long lifetime as a Trojan (Pilat-Lohinger and Dvorack 1999, Dvorack and
Tsiganis 2000). 

In order to test the applicability of $LCE$, we integrated the equations
of motion of the orbits of $32$ Trojan asteroids, using Mercury 6 (Chambers 1999),
taking into account the gravitational interactions of the Sun, Jupiter and
Saturn. The numerical integration of the orbit is terminated
when the asteroid ceases to co-orbit with Jupiter, defined to be when the
semi-major axis moves $0.4 AU$ away from the semi-major axis of the planet. All
the integrations were performed for at least $3.06 Gy$. The objects selected
for the dynamical study belong to our sample of available visible
spectra and include all the $24$ objects observed by us. 

Our results are summarised in table~\ref{tab:dataLCE}. We conclude that 
all objects with $LCE<0.53 \times 1/(10^5 yr)$ are macroscopically
stable and most objects with $LCE>0.7 \times 1/(10^5 yr)$ are unstable, but
a few ($~22\%$) may have found islands of stability, most of them with
exponents in the range $0.7 \times 1/(10^5 yr)<LCE<1.0 \times 1/(10^5 yr)$,
which agrees well with previous stability studies (Tsiganis \etal 2005).

\begin{table}[bp]
\begin{tabular}{|l|c|c|c|}\hline
$LCE$ & Number of & Percentage with lifetimes \\
interval    & objects &   $\geq 3 Gyr$\\
$(\times 1/(10^5 yr)$ &     &   \\ \hline
$LCE$$<$0.53  &  10 Objects &  $100\%$ \\
$LCE$$>$0.77 & 22 Objects  & $22\%$ \\    \hline
\end{tabular}
\caption{$LCE$ and macroscopic stability in our sample of Trojan asteroids. 
 It must be noted that there are no asteroids in our sample with
 $LCE$-values between $0.53 \times 1/(10^5 yr)$ and $0.77\times 1/(10^5 yr)$. }
\label{tab:dataLCE}
\end{table}

We conclude that, for our purposes, the use of $LCE$ is adequate
to determine statistically the macroscopic stability of the
Trojan asteroids, but we must bear in mind that there is a
small `contamination' of the population deemed by the $LCE$ to be 
unstable that are in fact stable.

\section{Visible Spectra of Trojan asteroids } 
\label{sec:vs}

To obtain as wide a data-base as possible on the spectra of Trojan
asteroids, we have used both data from our own observational program and
published data from a number of other sources.   

\subsection{New Observations} 

To obtain results with the best possible statistical significance, we performed our observations at a wavelength where the existing data is the
most abundant, which is, naturally, the visible range. Our sample, as
mentioned earlier, deliberately consists mainly of transitional objects. All
objects in the dynamical study were included in the observing 
sample.

Visible spectra were obtained with the 3.5m New Technology Telescope 
(NTT), at ESO La Silla (Chile), the 4.2m William Herschel (WHT) and the 
2.5 m Nordic Optical Telescope (NOT) both at the ``Roque de los Muchachos''
Observatory (ORM, La Palma, Spain).

At NTT the RILD arm of EMMI with the grism\#7 (150gr/mm) was used, covering
the $5200-9300 \AA$ spectral range, with a dispersion of $3.6 \AA$/pix.
Spectra were taken through a 5 arcsec wide slit. Observational circumstances
for NTT observations are shown in tables~\ref{tab:dataESO}.

At the WHT the red red arm of ISIS spectrograph with the R158R grating 
(158gr/mm) centred at $7500\AA$ and a second order blocking filter that cut at 
0.495 $\mu$m was used, covering the $5000-9500\AA$ spectral range, 
with a dispersion of $1.63\AA$/pixel. Spectra were 
taken through a 5arcsec wide slit. Observational circumstances for WHT
observations are shown in table~\ref{tab:dataWHT}.

At the NOT the ALFOSC (Andalucia Faint Object Spectrograph and Camera)
with a grism disperser  \#4 (300gr/mm) and GG475 second order blocking 
filter was used, covering the $4800-9100 \AA$ spectral range, 
with a dispersion of $3\AA$/pixel.  Spectra were 
taken through a 1.3 arcsec wide slit. Observational circumstances for NOT
observations are shown in table~\ref{tab:dataNOT}.

In all telescopes the slit was oriented in the parallactic angle, and the 
tracking was at the asteroid proper motion.
 
\begin{figure}[bp]
\centerline{\includegraphics[width=\textwidth,height=\textheight]{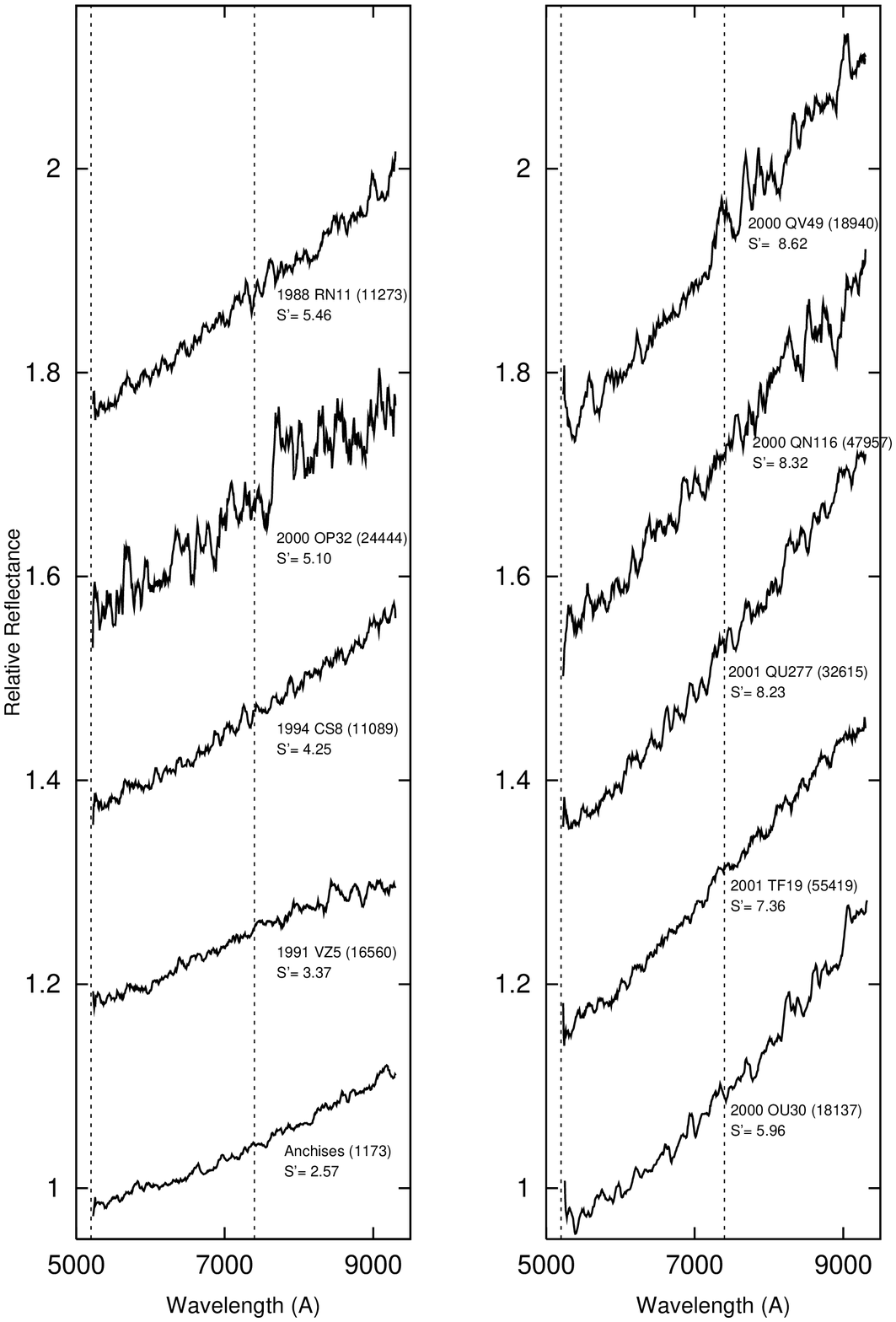}}
\caption{Spectra of the  10 Trojan asteroids observed at NTT-ESO. The relative
reflectance is shifted vertically by and additive constant for clarity. The vertical
dotted lines indicate the range over which the linear fit has been made.}
\label{fig:specESO1}
\end{figure}

\begin{figure}[bp]
\centerline{\includegraphics[width=\textwidth,height=\textheight]{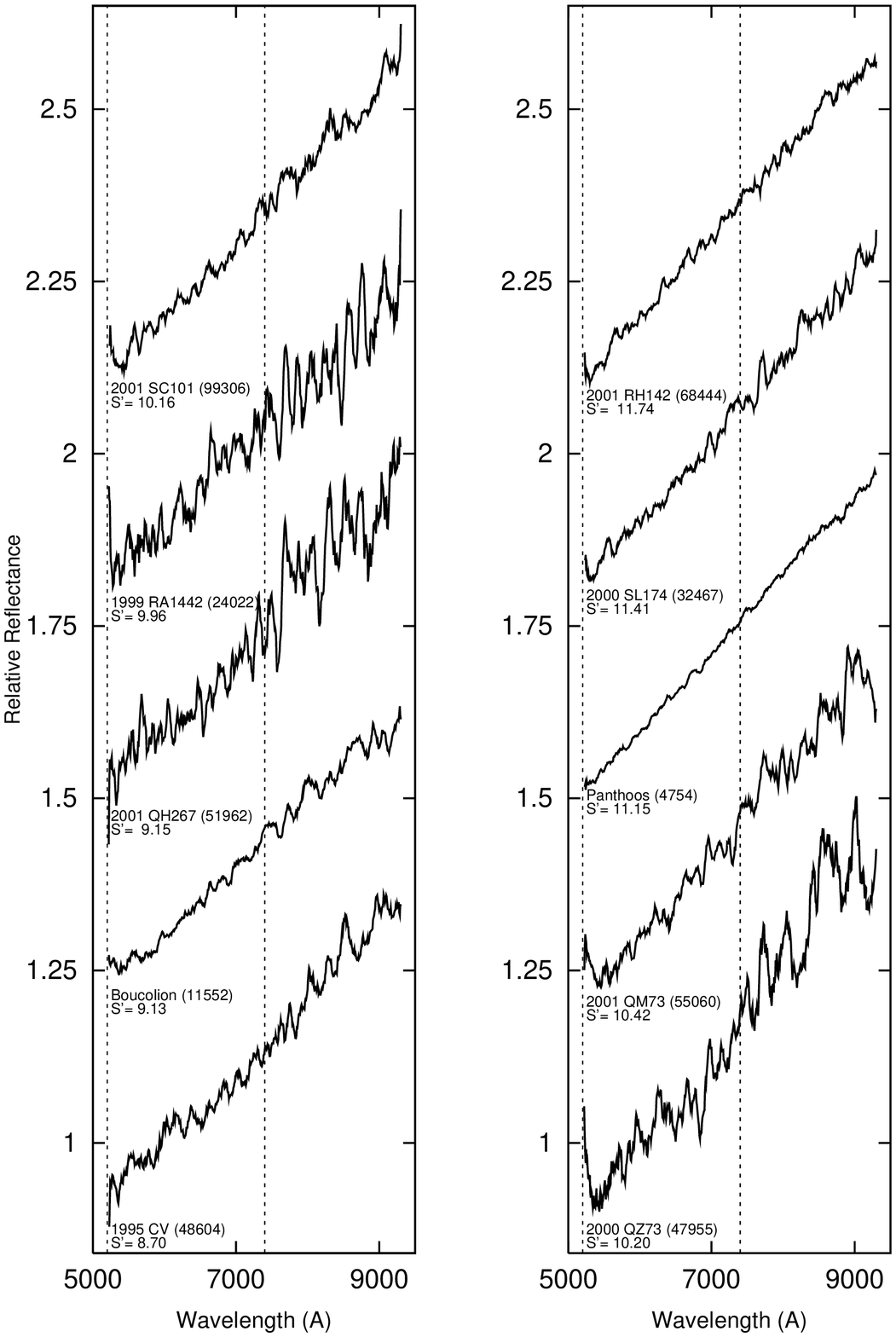}}
\caption{Spectra of the 10 Trojan asteroids observed at NTT-ESO (cont.). The
relative
reflectance is again shifted by and additive constant for clarity. The vertical
dotted lines indicate the range over which the linear fit has been made.}
\label{fig:specESO2}
\end{figure}

\begin{figure}[bp]
\centerline{\includegraphics[width=10cm,height=\textheight]{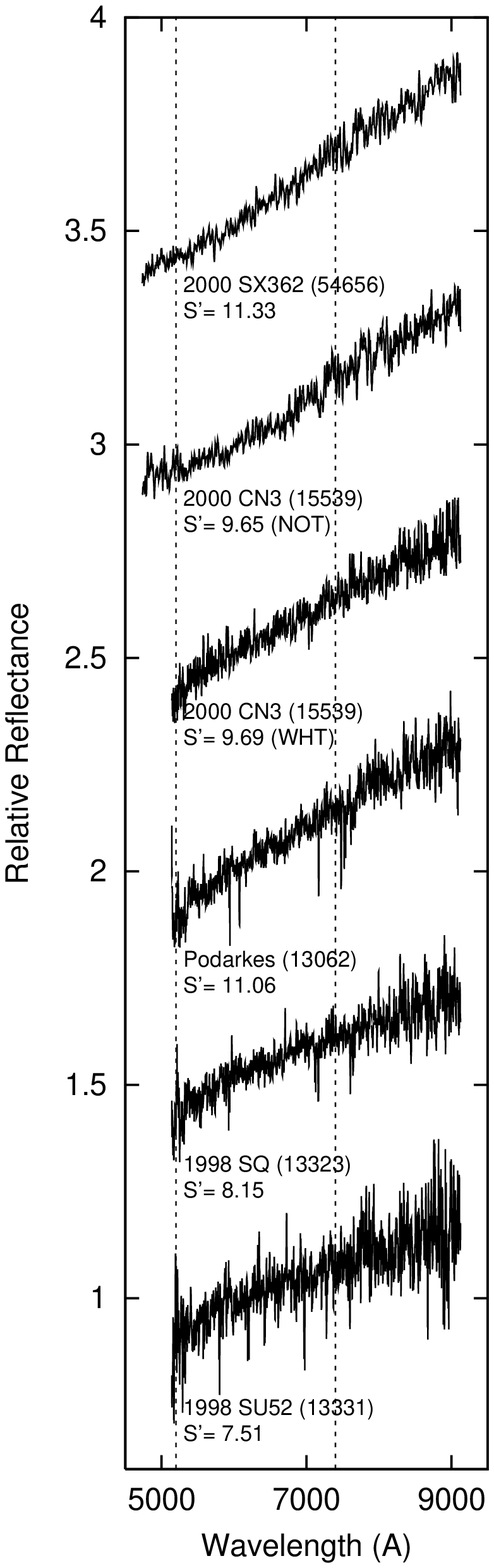}}
\caption{Spectra of the 5 Trojan asteroids observed at WHT-ING and NOT. The
relatively
reflectance is again shifted by and additive constant for clarity. The vertical
dotted lines indicate the range over which the linear fit has been made.}
\label{fig:specNOTWHT}
\end{figure}

Data reduction was carried out in the standard way using standard IRAF procedures. Images were
over-scan and bias corrected, and flat-field corrected using lamp flats. The
two-dimensional spectra were extracted, sky background subtracted, and
collapsed to one dimension. The wavelength calibration was done using
Helium, Neon and Argon lamps. The reflectance spectra were obtained by
dividing the spectra of the asteroids 
by the spectra of Solar-analogue star Hyades 64 and G2 stars Landolt
98-978 and Landolt 102-1081 (Landolt, 1992), 
observed during the same night at airmasses
similar to that of the asteroids. Reflectance spectra, normalised at 0.6
$\mu$m, are shown in figures~\ref{fig:specESO1}, \ref{fig:specESO2}
and~\ref{fig:specNOTWHT}.  

We characterise the color of the surfaces by the reflectivity gradient,
$S'$, in the wavelength interval [$5100$\AA\ --\ $7400$\AA], where: \[S' =
{1}/{S(6000 \AA)}\ \frac{\Delta S}{\Delta \lambda},\] $S$ is the normalised
reflectivity with respect to a Solar-analogue star and $\lambda$, the
wavelength. We have chosen a wavelength range where we have data from all
the available observations. 

The value of $S'$, computed in that wavelength range, gives a good
estimation of the nature of the surface. Most Trojans are quite red objects
and have been classified as $D$-types or $P$-types (see for example Bendjoya
\etal 2004), which means that their spectra are featureless and linear, with
a positive slope. 

The obtained values of $S'$ for our own group of targets are given in
table~\ref{tab:Sp}, where we also give the corresponding value
of LCE and the absolute magnitudes, $H$, 
taken from the {\it
AstDys} database\footnote{http://hamilton.dm.unipi.it/cgi-bin/astdys/astibo}.

\begin{table}[bp]
\begin{tabular}{|l|c|c|c|c|c|}\hline
Designation & Airmass & Date & UT & exp. time & N \\ \hline
 (1173)  Anchises   & 1.42    & 2005-01-08 & 06:51:44.502 &  300 & 3 \\
 (4754)  Panthoos    & 1.33  & 2005-01-09 & 08:25:26.240 &  400 & 2 \\
 (11089) 1994 CS8   & 1.57  & 2005-01-08 & 04:40:35.215 &  300 & 4 \\
 (11273) 1988 RN11  & 1.58  & 2005-01-09 & 02:44:13.597 &  600 & 3 \\
 (11552) Boucolion   & 1.39 & 2005-01-10 & 08:14:30.182 &  900 & 2 \\
 (16560) 1991 VZ5   & 1.68  & 2005-01-09 & 04:38:24.950 &  300 & 3 \\
 (18137) 2000 OU30  & 1.42  & 2005-01-08 & 03:32:46.758 &  600 & 3 \\
 (18940) 2000 QV49  & 1.51  & 2005-01-08 & 05:46:04.529 &  900 & 3 \\
 (24022) 1999 RA144  & 1.18  & 2005-01-10 & 07:01:32.173 &  1200 & 3 \\
 (24444) 2000 OP32  & 1.76  & 2005-01-10 & 03:06:04.229 &  900 & 2 \\
 (32467) 2000 SL174  & 1.01  & 2005-01-08 & 08:05:26.299 &  600 & 3 \\
 (32615) 2001 QU277 & 1.58  & 2005-01-08 & 02:42:36.883 &  300 & 3 \\
 (47955) 2000 QZ73   & 1.44  & 2005-01-10 & 05:34:28.543 &  1200 & 3 \\
 (47957) 2000 QN116 & 1.50  & 2005-01-09 & 06:24:33.763 &  600 & 3 \\
 (48604) 1995 CV    & 1.46  & 2005-01-10 & 02:07:22.901 &  1200 & 2 \\
          ``        & 1.50  & 2005-01-09 & 01:37:15.375 &  900 & 3 \\
 (51962) 2001 QH267  & 1.84  & 2005-01-10 & 03:50:52.596 &  900 & 2 \\
 (55060) 2001 QM73   & 1.74  & 2005-01-10 & 04:44:00.799 &  1200 & 2 \\
 (55419) 2001 TF19  & 1.65  & 2005-01-09 & 05:10:48.370 &  500 & 3 \\
 (68444) 2001 RH142  & 1.26  & 2005-01-09 & 03:48:49.472 &  400 & 3 \\
 (99306) 2001 SC101  & 1.56  & 2005-01-08 & 01:31:47.437 &  900 & 3 \\  \hline
\end{tabular}
\caption{Observational circumstances of the objects observed at NTT-ESO. 
The exposure time corresponds to individual images. N is the total number of
images taken. }
\label{tab:dataESO}
\end{table}

\begin{table}[bp]
\begin{tabular}{|l|c|c|c|c|c|}\hline
Designation & Airmass & Date & UT & exp. time (sec.) & N \\ \hline
(13062) Podarkes  & 1.48  &   2004-06-14  &     00:06  &    600 & 3 \\
(13331) 1998 SU52 & 1.38  &   2004-06-13  &     21:53  &    900 & 3 \\
(13323) 1998 SQ   & 1.39  &   2004-06-13  &     22:48  &    600 & 4  \\
(15539) 2000 CN3  & 1.08  &   2004-06-13  &     23:49  &    400 & 3 \\  \hline
\end{tabular}
\caption{Observational circumstances of the objects observed at WHT-ING.
The exposure time corresponds to individual images. N is the total number of
images taken.}
\label{tab:dataWHT}
\end{table}

\begin{table}[bp]
\begin{tabular}{|l|c|c|c|c|c|}\hline
Designation & Airmass & Date & UT & exp. time (sec.) & N \\ \hline
(15539) 2000 CN3    &    1.18 &   2004-08-31  &    20:06:20  &    600 & 3 \\
(54656) 2000 SX362  &    1.28 &   2004-08-31  &    19:20:02  &    600 & 3 \\
\hline
\end{tabular}
\caption{Observational circumstances of the objects observed at NOT.
The exposure time corresponds to individual images. N is the total number of
images taken.}
\label{tab:dataNOT}
\end{table}

\begin{table}[bp]
\begin{tabular}{|l|c|c|c|c|c|}\hline
Designation & $LCE$ ($ \times 1/(10^5 yr)$) & $S'$ (\% / 1000$\AA$)&
$\sigma_{S'}$ (\% / 1000$\AA$) & Swarm & H \\ \hline
 (1173)  Anchises     &       2.41    &     2.57    &     0.03  &  L5   &  8.89   \\
 (4754)  Panthoos     &       0.77     &   11.15     &    0.03  &  L5   &  10.1    \\ 
 (11089) 1994 CS8     &       1.78    &     4.25    &     0.05  &  L5   &  10.3   \\  
 (11273) 1988 RN11    &       2.15    &     5.46    &     0.06  &  L5   &  11.6   \\  
 (11552) Boucolion    &       1.10     &    9.13     &    0.09  &  L5   &  10.6   \\  
 (13062) Podarkes       & 2.61         &  11.06 &  0.12         &  L4   &  11.1   \\
 (13323) 1998 SQ        & 4.04         &   8.15 &  0.12         &  L4   &  10.7   \\
 (13331) 1998 SU52      & 6.91         &   7.53 &  0.20         &  L4   &  11.2    \\
 (15539) 2000 CN3 - NOT & 3.01         &   9.65 &  0.13         &  L4   &   9.9    \\
 (15539) 2000 CN3 - WHT & 3.01         &   9.69 &  0.10         &  L4   &   9.9    \\
 (16560) 1991 VZ5     &       1.19    &     3.37    &     0.04  &  L5   &  10.9   \\  
 (18137) 2000 OU30    &       4.00    &     5.96    &     0.08  &  L5   &  11.4   \\  
 (18940) 2000 QV49    &       1.99    &     8.62    &     0.13  &  L5   &  11.4   \\  
 (24022) 1999 RA144   &       2.46    &     9.96    &     0.22  &  L5   &  11.8   \\  
 (24444) 2000 OP32    &       3.25    &     5.11    &     0.15  &  L5   &  11.2   \\  
 (32467) 2000 SL174   &       2.66    &    11.41    &     0.09  &  L5   &  11.7     \\
 (32615) 2001 QU277   &       1.02    &     8.23    &     0.09  &  L5   &  10.5   \\  
 (47955) 2000 QZ73    &       1.58    &    10.20    &     0.24  &  L5   &  11.4   \\  
 (47957) 2000 QN116   &       1.42    &     8.32    &     0.10  &  L5   &  12.6   \\  
 (48604) 1995 CV      &       1.75    &     8.70    &     0.12  &  L5   &  10.8   \\  
 (51962) 2001 QH267   &       0.83    &     9.15    &     0.21  &  L5   &  10.4   \\  
 (54656) 2000 SX362     & 0.13         &  11.33 &  0.10         &  L5   &  10.7   \\ 
 (55060) 2001 QM73    &       0.97    &    10.42    &     0.16  &  L5   &  11.6   \\ 
 (55419) 2001 TF19    &       1.15    &     7.36    &     0.06  &  L5   &  10.7   \\
 (68444) 2001 RH142   &       1.29    &    11.74    &     0.07  &  L5   &  11.2    \\ 
 (99306) 2001 SC101   &       1.53    &    10.16    &     0.12  &  L5   & 12.6   \\   \hline
\end{tabular}
\caption{Dynamical stability and reflectivity gradient in our
sample: values of the $LCE$ and $S'$ for the Trojan asteroids observed by
ourselves. The corresponding Lagrangian point (Swarm) and the Absolute magnitude, H,
is also indicated. }
\label{tab:Sp}
\end{table}

\subsection{Spectral information from other published databases}

\label{sec:noowndata}

The data used in this work were taken from : Jewitt \& Luu (1990) (32
observations Ref id: 1), SMASSII (Bus, 1999, 5 observations, Ref id: 2),
Bendjoya \etal (2004) (34 observations, Ref id: 3), Lazzaro \etal (2004) (10
observations, Ref id: 4), Dotto \etal (2006 Ref id: 5)\footnote{Only a
subset of the spectra contained in this article is available to us.} (5
observations). The data from Fornasier et al (2007) is not 
available to us. Their sample is composed of 47 objects, all family members,
whose surface properties may not statistically mimic those of the whole
Trojan population so they may have introduced an unwanted bias.

We have computed the value of $S'$ by a linear fit in the [$5100$\AA\ --\
$7400$\AA] wavelength range for the 
spectra found in the databases of Bus (1999) and Lazzaro \etal (2004). 
The spectra from Jewitt and Luu (1990) was scanned to compute
the value of $S'$ in that same interval. For the article of Bendjoya \etal
(2004) we used the quoted value of $S'$, since the interval is very similar
to the one chosen by us. All the correspondent values are listed in
tables~\ref{tab:Spout1}, \ref{tab:Spout2} and~\ref{tab:Spout3}. 

\begin{table}[bp]
\begin{tabular}{|l|c|c|c|c|c|c|}\hline
Desig. & $LCE$ ($ \times 1/(10^5 yr)$) & $S'$ (\% / 1000$\AA$)
& $\sigma_{S'}$ (\% / 1000$\AA$) & H (mag) & Swarm & Ref. id.\\ \hline
          588  &   0.02  &   10.50   &   0.20   &   8.7    &   L4   &    3 \\
          624  &   0.14  &    9.03   &   0.05   &   7.5    &   L4   &    1 \\
          659  &   0.20  &    1.66   &   0.16   &   9.0    &   L4   &    1 \\
          884  &   0.18  &    8.79   &   0.10   &   8.8    &   L5   &    1 \\
          911  &   0.15  &   13.60   &   0.10   &   7.9    &   L4   &    3 \\
         1143  &   0.11  &   11.26   &   0.33   &   7.9    &   L4   &    2 \\
         1172  &   0.10  &   12.30   &   0.10   &   8.3    &   L5   &    3 \\
         1172  &   0.10  &    8.11   &   0.09   &   8.33    &   L5   &    1 \\
         1173  &   2.41  &    2.72   &   0.08   &   8.9    &   L5   &    1 \\
         1208  &   0.08  &    1.81   &   0.14   &   9.0    &   L5   &    1 \\
         1437  &   0.74  &    2.41   &   0.07   &   8.3    &   L4   &    1 \\
         1583  &   0.06  &    8.56   &   0.09   &   8.6    &   L4   &    1 \\
         1647  &   0.07  &    8.86   &   0.35   &  10.3    &   L4   &    1 \\
         1749  &   0.19  &    8.62   &   0.10   &   9.2    &   L4   &    1 \\
         1749  &   0.19  &    9.01   &   0.26   &   9.2    &   L4   &    2 \\
         1867  &   0.48  &   10.91   &   0.09   &   8.6    &   L5   &    1 \\
         1868  &   1.97  &    6.70   &   0.10   &   9.3    &   L4   &    3 \\
         1870  &   0.03  &   20.34   &   0.48   &  10.5    &   L5   &    1 \\
         1871  &   0.51  &   12.38   &   0.89   &  11.0    &   L5   &    1 \\
         1872  &   0.20  &   25.42   &   1.15   &  11.2    &   L5   &    1 \\
         1873  &   0.25  &   10.30   &   0.36   &  10.5    &   L5   &    1 \\
         2207  &   0.25  &    9.99   &   0.12   &   9.0    &   L5   &    1 \\
         2223  &   0.21  &   15.90   &   0.10   &   9.4    &   L5   &    3 \\
         2241  &   0.28  &    8.43   &   0.11   &   8.6    &   L5   &    1 \\
         2260  &   0.06  &   11.32   &   0.26   &   9.3    &   L4   &    1 \\
         2357  &   0.03  &    9.27   &   0.12   &   8.9    &   L5   &    1 \\
         2363  &   0.01  &    8.32   &   0.21   &   9.1    &   L5   &    1 \\
         2456  &   0.29  &    7.79   &   0.11   &   9.6    &   L4   &    1 \\ 
         2674  &   0.12  &   10.54   &   0.15   &   9.0    &   L5   &    1 \\ 
	 \hline
\end{tabular} 
\caption{Data values for the Trojan asteroids taken from
external databases I.}
\label{tab:Spout1}
\end{table}

\begin{table}[bp]
\begin{tabular}{|l|c|c|c|c|c|c|}\hline
Desig. & $LCE$ ($\times 1/(10^5 yr)$) & $S'$ (\% / 1000$\AA$)
& $\sigma_{S'}$ (\% / 1000$\AA$) & H (mag) & Swarm & Ref. id.\\ \hline
2759  &   0.08  &    9.90   &   0.16   &   9.8    &   L4   &    1 \\
2797  &   0.70  &    8.28   &   0.07  &   8.4    &   L4   &    1 \\
         2895  &   0.05  &    0.10   &   0.60   &   9.3    &   L5   &    3 \\
         2920  &   0.28  &   10.17  &   0.21   &   8.8    &   L4   &    2 \\
         2920  &   0.28  &    9.80   &   0.08   &   8.8    &   L4   &    1 \\
         3063  &   0.21  &    7.95   &   0.05  &   8.6    &   L4   &    4 \\
         3240  &   0.03  &    7.68   &   0.31   &  10.0    &   L5   &    1 \\
         3317  &   0.05  &   10.47   &   0.04   &   8.3    &   L5   &    3 \\
         3317  &   0.05  &    7.54   &   0.12   &   8.3    &   L5   &    2 \\
         3391  &   0.38  &    4.69   &   0.31   &  10.3    &   L4   &    1 \\
         3451  &   0.90  &    1.89   &   0.10   &   8.1    &   L5   &    2 \\
         3451  &   0.90  &    2.24   &   0.09   &   8.1    &   L5   &    1 \\
         3708  &   0.03  &    9.40   &   0.10   &   9.3    &   L5   &    3 \\
         3709  &   0.04  &   10.95   &   0.07   &   9.0    &   L4   &    4 \\
         3709  &   0.04  &    8.40   &   0.10   &   9.0    &   L4   &    3 \\
         3793  &   0.01  &    3.66  &   0.13   &   8.8   &   L4   &    1 \\
         3793  &   0.01  &    7.44   &   0.07   &   8.8    &   L4   &    4 \\
         4035  &   0.11  &   15.45   &   0.05   &   9.3    &   L4   &    5 \\
         4035  &   0.11  &    8.80   &   0.10   &   9.3    &   L4   &    3 \\
         4060  &   0.12  &    5.50   &   0.10   &   8.9    &   L4   &    3 \\
         4063  &   0.15  &    6.64   &   0.04   &   8.6    &   L4   &    4 \\
         4063  &   0.15  &    8.40   &   0.10   &   8.6    &   L4   &    3 \\
         4068  &   0.42  &   10.27   &   0.04   &   9.4    &   L4   &    4 \\
         4068  &   0.42  &   10.30   &   0.14   &   9.4    &   L4   &    1 \\
	 4068  &   0.42  &   14.40   &   0.10   &   9.4    &   L4   &    3 \\ 
4138  &   0.07  &    2.00   &   0.20   &   9.8  &   L4   &    3 \\
         4348  &   0.13  &    3.98   &   0.04   &   9.2    &   L5   &    3 \\
         4489  &   0.53  &    7.90  &   0.09   &   9.0    &   L4   &    4 \\
	          4715  &   0.15  &   13.50   &   0.20   &   9.3    &   L5 & 3 \\ \hline
\end{tabular}
\caption{Data values for the Trojan asteroids taken from
external databases II. }
\label{tab:Spout2}
\end{table}

\begin{table}[bp]
\begin{tabular}{|l|c|c|c|c|c|c|}\hline
 Desig. & $LCE$ ($ \times 1/(10^5 yr$) & $S'$ (\% / 1000$\AA$)
 & $\sigma_{S'}$ (\% / 1000$\AA$) & H (mag) & Swarm & Ref. id.\\ \hline
         4792  &   0.10  &   14.00   &   0.20   &  10.0    &   L5   &    3 \\
         4833  &   0.31  &   10.04   &   0.09   &   9.1    &   L4   &    4 \\
         4833  &   0.31  &   11.20   &   0.10   &   9.1    &   L4   &    3 \\
         4834  &   0.12  &    9.80   &   0.10   &   9.2    &   L4   &    3 \\
         4835  &   0.16  &    5.00   &   0.20   &   9.8    &   L4   &    3 \\
         4835  &   0.16  &    7.96   &   0.07   &   9.8    &   L4   &    4 \\
         4836  &   0.16  &    7.20   &   0.10   &   9.5    &   L4   &    3 \\
         4902  &   0.05  &    7.05   &   0.07   &   9.6    &   L4   &    4 \\
         5025  &   0.07  &   13.00   &   0.30   &   9.8    &   L4   &    3 \\
         5027  &   0.17  &    9.29   &   0.32   &   9.4    &   L4   &    1 \\
         5028  &   0.15  &   13.57   &   0.39   &   9.9    &   L4   &    1 \\
         5126  &   0.08  &    0.80   &   0.10   &  10.1    &   L4   &    3 \\
         5254  &   0.05  &   10.47   &   0.04   &   8.8    &   L4   &    3 \\
         5258  &   0.18  &    6.50   &   0.30   &  10.0    &   L4   &    3 \\
         5264  &   0.01  &   11.50   &   0.20   &   9.5    &   L4   &    3 \\
         5264  &   0.01  &    8.77   &   0.09   &   9.5    &   L4   &    4 \\
         5283  &   0.07  &   -8.90   &   0.10   &   9.3    &   L4   &    3 \\
         5285  &   0.31  &    5.70   &   0.10   &   9.8    &   L4   &    3 \\
         5511  &   0.13  &   13.00   &   0.10   &   9.6    &   L5   &    3 \\
         5648  &   0.33  &    4.90   &   0.40   &   9.0    &   L5   &    3 \\
         6090  &   0.32  &   11.70   &   0.20   &   9.4    &   L4   &    3 \\
         7152  &   0.07  &    3.30   &   0.20   &   9.9    &   L4   &    3 \\
         7352  &   0.47  &    9.80   &   0.20   &   9.0    &   L5   &    3 \\
         7641  &   0.02  &   -0.80   &   0.20   &   9.3    &   L4   &    3 \\
        11351  &   0.40  &   10.92   &   0.13   &  10.5    &   L4   &    5 \\
        12921  &   0.08  &    3.24   &   0.08   &  10.7    &   L4   &    5 \\
        20738  &   0.06  &    9.25   &   0.15   &  11.4    &   L4   &    5 \\
        24341  &   0.44  &   10.94   &   0.09   &  11.5    &   L4   &    5 \\ \hline
\end{tabular}
\caption{Data values for the Trojan asteroids taken from
external databases III.}
\label{tab:Spout3}
\end{table}

\subsection{Differences between the various data sets}

In all we have a total of $112$ observations corresponding to $96$ different
objects. There are $15$ objects that have $S'$ determined from two sources 
observations and two from three sources.

In table~\ref{tab:1} we have listed the objects that have repeated observations.
The median value of the differences in the values of $S'$ for the repeated
observations, $<\Delta$S'$>_{med}$, is: 
\[<\Delta S'>_{med} \sim 2.5.\] 
\begin{table}[bp]
\begin{tabular}{|l|c|c|c|}\hline
   Designation   &      $|\Delta S'|$ &    \multicolumn{2}{c|}{Between} \\ \hline  
   1172 &    4.19    & Jewitt  &   Bendoya \\
   1173 &   0.14    & Jewitt  &   NTT-ESO      \\
   1749 &   0.39    & Jewitt  &   SMASSII      \\
   2920 &   0.37    & Jewitt  &   SMASSII      \\
   3317 &    2.93    & Bendoya &   SMASSII      \\
   3451 &   0.35    & SMASSII  &   Jewitt    \\
   3709 &    2.55    & Bendoya &   Lazzaro  \\
   3793 &    3.77    & Jewitt  &   Lazzaro  \\
   4035 &    6.65    & Bendoya &   Dotto   \\
   4063 &    1.76    & Bendoya &   Lazzaro  \\
   4068 &    4.13    & Bendoya &   Lazzaro \\
   4068 &    4.10    & Bendoya  &   Jewitt  \\
   4068 &   0.04 & Jewitt &    Lazzaro  \\
   4833 &    1.16    & Bendoya &   Lazzaro  \\
   4835 &    2.96    & Bendoya &   Lazzaro  \\
   5264 &    2.73    & Bendoya &   Lazzaro  \\
  15539 &   0.04 & WHT-ING    &    NOT      \\ \hline
\end{tabular}\\
\caption{Magnitudes of the differences in $S'$, $|\Delta S'|$,
for objects repeated in the available sample.}
\label{tab:1}
\end{table}
Based on this, we adopted a value for the typical observation error of $S'$
as $<\Delta S' >_{med}$. Those observations with values of $S'_d$, such that
(with $|S'_d - <S'>_{med}| > 3 \times <\Delta S'>_{med}$) will be discarded
from the analyses. Therefore objects with values of $S' \lessapprox 0$ and $S'
\gtrapprox 15$ will not be considered. We exclude from our sample the
following objects: 5283 ($S'=-8.9$, Ref id.:3), 1870 ($S'=20.34$, Ref id.:1)
and 1872 ($S'=25.42$, Ref id.:1). Therefore the total number of objects of
the sample is $93$. For objects with more than one value of $S'$ originating
in different sources, we use their mean value.

We also note that the median of differences is much smaller if the objects
from Bendoya \etal (2004) are excluded, giving: $<\Delta S' >_{med} = 0.35$.

\section{Correlations between various properties}
\label{sec:corr}

\subsection{Size distributions of the resident and the transitional groups}
\label{sec:corrsize}

In this section we compare the size distributions of the resident and the
transitional Trojans. If the unstable orbits are mainly fed by the smallest
by-products of collisions, the size distributions of the stable and the
unstable groups will be different, but if the only mechanism by which a
Trojan asteroid reaches an unstable orbit is dynamical, then both the size
distributions will be similar.

\begin{figure}[bp]
\centerline{\includegraphics[width=\textwidth,height=\textheight]{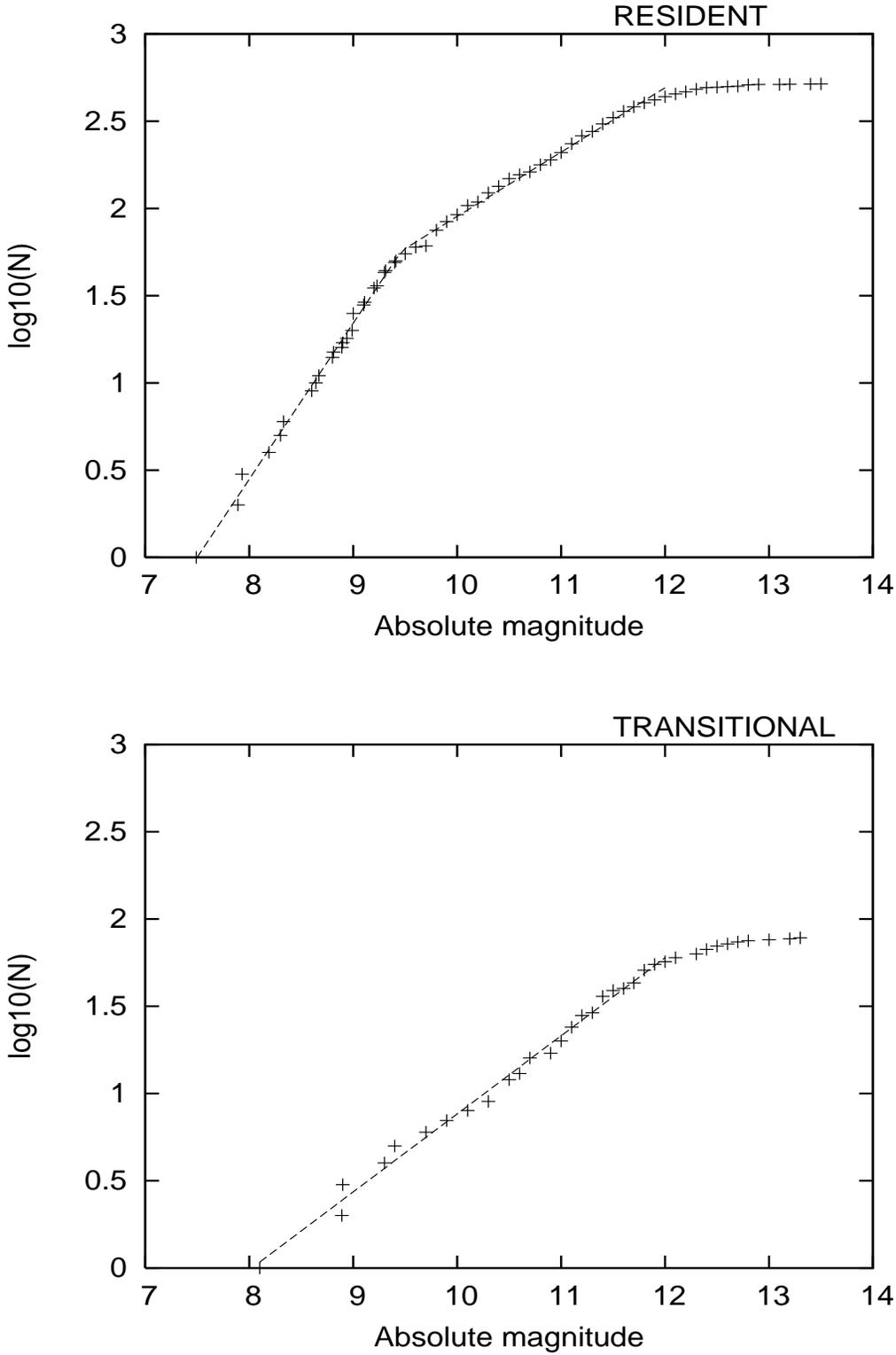}}
\caption{Absolute magnitude distributions of the `Resident' and the
`Transitional' groups. The straight lines correspond to the best linear fits.}
\label{fig:1}
\end{figure}

We assume that the absolute magnitude, $H$, is a reliable indicator of the
size of the asteroids. The absolute magnitudes, $H$, of the asteroids are
taken from {\it AstDys}.  The
$H$-distribution of both groups is shown in figure~\ref{fig:1}. We
considered only multi-opposition objects with $H<12$, since it is apparent
from figure~\ref{fig:1} that for fainter objects the sample has additional
bias selection effects. 

We have divided the numbered Trojan asteroids into 2 groups, according to the
value of $LCE$, such that the {\it resident} Trojans have $LCE_{res} \le
0.53\ \times 1/(10^5 yr)$ and the {\it transitional} ones, $LCE_{trans} > 0.53\
1/(10^5 yr)$. According to this criteria, in the {\it AstDys} database we
have found $517$ resident and $121$
transitional multi-opposition Trojans.  

Realistic variations in the geometrical albedo, $p_V$, for values in the  
range observed in the Trojan populations, makes little difference to the 
value of the diameter. Assuming a mean geometrical albedo of $p_V=0.04$ 
(Fernandez et al. \etal 2003), the objects in our sample have diameters 
between $20 km$ and $260 km$ approximately. 

\begin{figure}[bp]
\centerline{\includegraphics{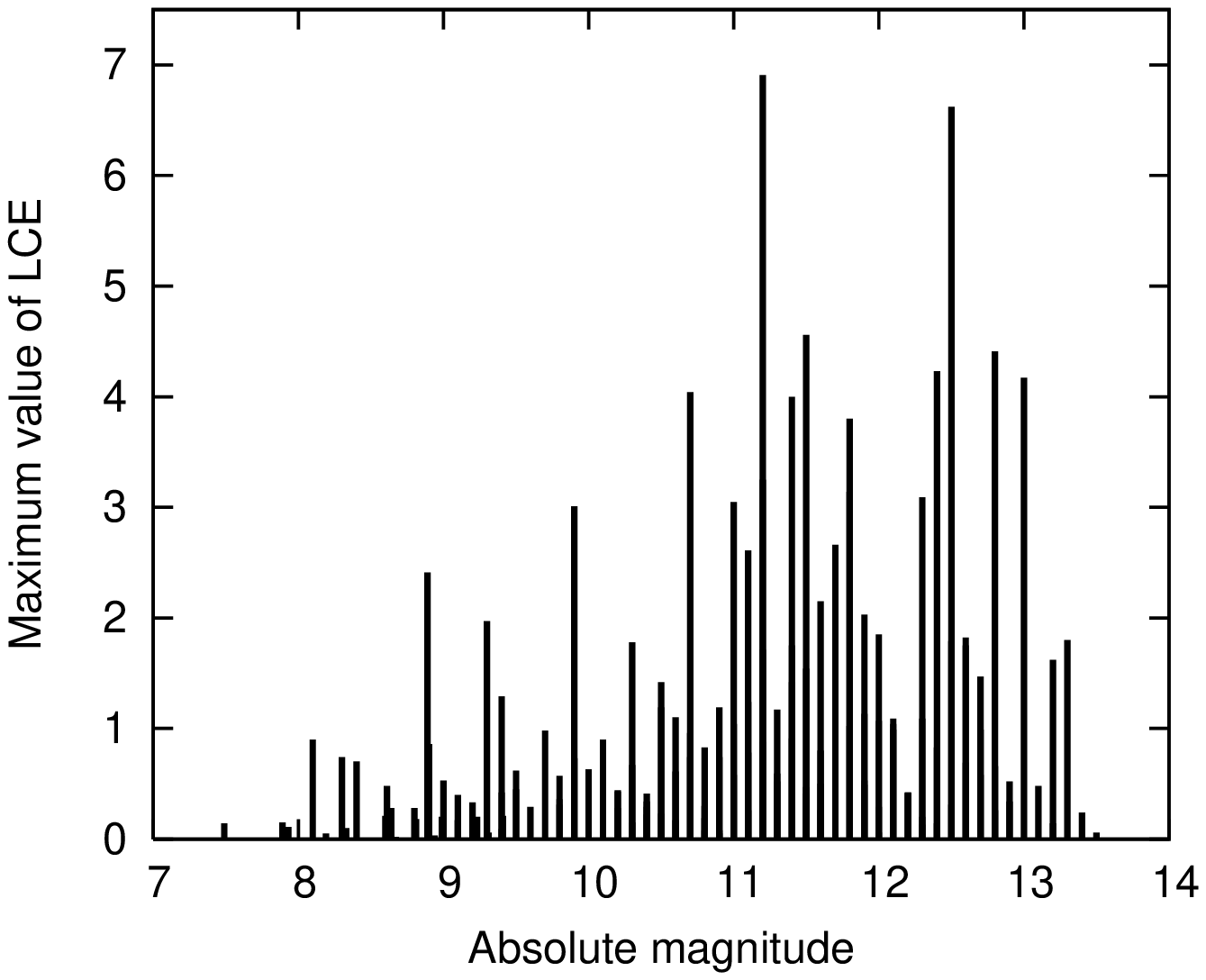}}
\caption{ The maximum value of $LCE$ (in units of $1/(10^5 yr)$) at a
particular value of $H$ for the numbered Trojans. It is apparent that the
more unstable orbits tend to be populated by smaller objects.}
\label{fig:HHLCE}
\end{figure}

As we can see from fig~\ref{fig:HHLCE}, there is a tendency for objects in
more unstable orbits to be smaller.

The absolute magnitude distributions of these groups, shown in
figure~\ref{fig:1}, are fit
with linear functions of the type: 
$$log_{10}(N) = b\ H + a.$$

\begin{figure}[bp]
\centerline{\includegraphics[width=\textwidth,height=20cm]{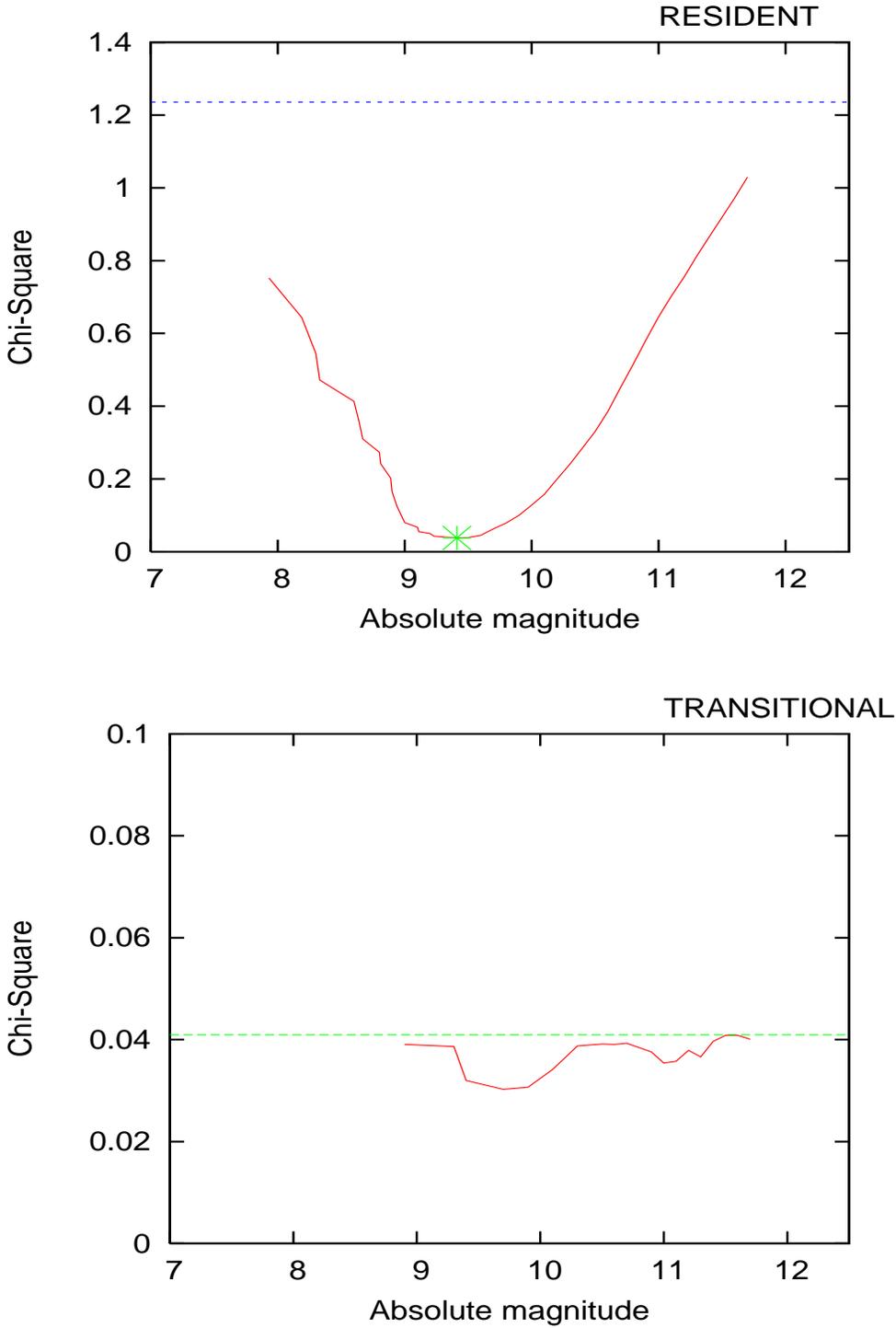}} 
\caption{The values of $\chi^2$ as a function of the location of the
nodal point, for each group. The value of $\chi^2$ corresponding to
a fit with a single linear function from beginning to end is indicated by an
horizontal line. The value of the nodal point that gives the minimum
$\chi^2$ for the resident Trojans, $H_{node} = 9.41$, is also marked.
The absolute magnitude distribution of the resident Trojans is clearly
bimodal while the one of the transitional ones is unimodal (notice the
change in the scale of the y-axis).}
\label{fig:2}
\end{figure}

For a bimodal distribution the fit is done between the absolute magnitude
distribution and $2$ linear-functions:
\begin{eqnarray}
log_{10}(N_1) = b_1\ H + a_1 & H<H_{n} \nonumber \\
log_{10}(N_2) = b_2\ H + a_2 & 12>H>H_{n} \nonumber \\
\end{eqnarray}
We also require that:
$$ a_1 + b_1\ H_{n} = a_2 + b_2 H_{n} .$$
We calculate the corresponding values of $\chi^2$ as a function of $H_{n}$.
The first and last values of $H_{n}$ are chosen to include $5$ data-points. 
The nodal point, $H_{node}$, corresponds to the value of $H_{n}$ that gives
the minimum value of $\chi^2$ (see figure~\ref{fig:2}).  
We find that the value of the nodal point for the resident Trojans is
$H_{node} = 9.41$.  

For the transitional Trojans, the values of $\chi^2$ obtained with the
bimodal fit are not significantly different from the one of a single linear
fit (see also figure~\ref{fig:2}).  

The value of slopes corresponding to the best fits are shown in 
table~\ref{tab:22}. It is apparent that the absolute magnitude distribution
of `small' resident Trojans is similar to the one of the transitional ones.
Given that the distribution of albedos in the Trojans is rather narrow
(Fernandez \etal 2003), the size-distributions must be similar to the H
distribution, therefore it is plausible that the Transitional Trojans are
mainly collisional fragments. For the whole population Jewitt \etal 2000
also obtain a bimodal size-distribution, the nodal point is located at
visual absolute magnitudes $V \approx 10$, the index for the faint branch
is $0.4$ and the one for the bright branch is $1.12$.   

\begin{table}[bp]
\begin{tabular}{|l|c|c|}\hline
Group & $b$ & $\chi^2$ \\ \hline
Big resident ($H < 9.41$) &  $0.89$ &  $0.037$  \\
Small resident ($H \ge 9.41$) & $0.37$ &   \\ 
Transitional & $0.45$  &  $0.041$ \\ \hline
\end{tabular}\\
\caption{Slopes of the linear fits of the size distribution of the Resident
(Big and Small) and Transitional groups. Notice that the value of $\chi^2$
for the resident group corresponds to a single bimodal fit.}
\label{tab:22}
\end{table}

\subsection{Visual reflection spectra, absolute magnitude, dynamical
properties and albedos}

\begin{figure}
\includegraphics[width=\textwidth]{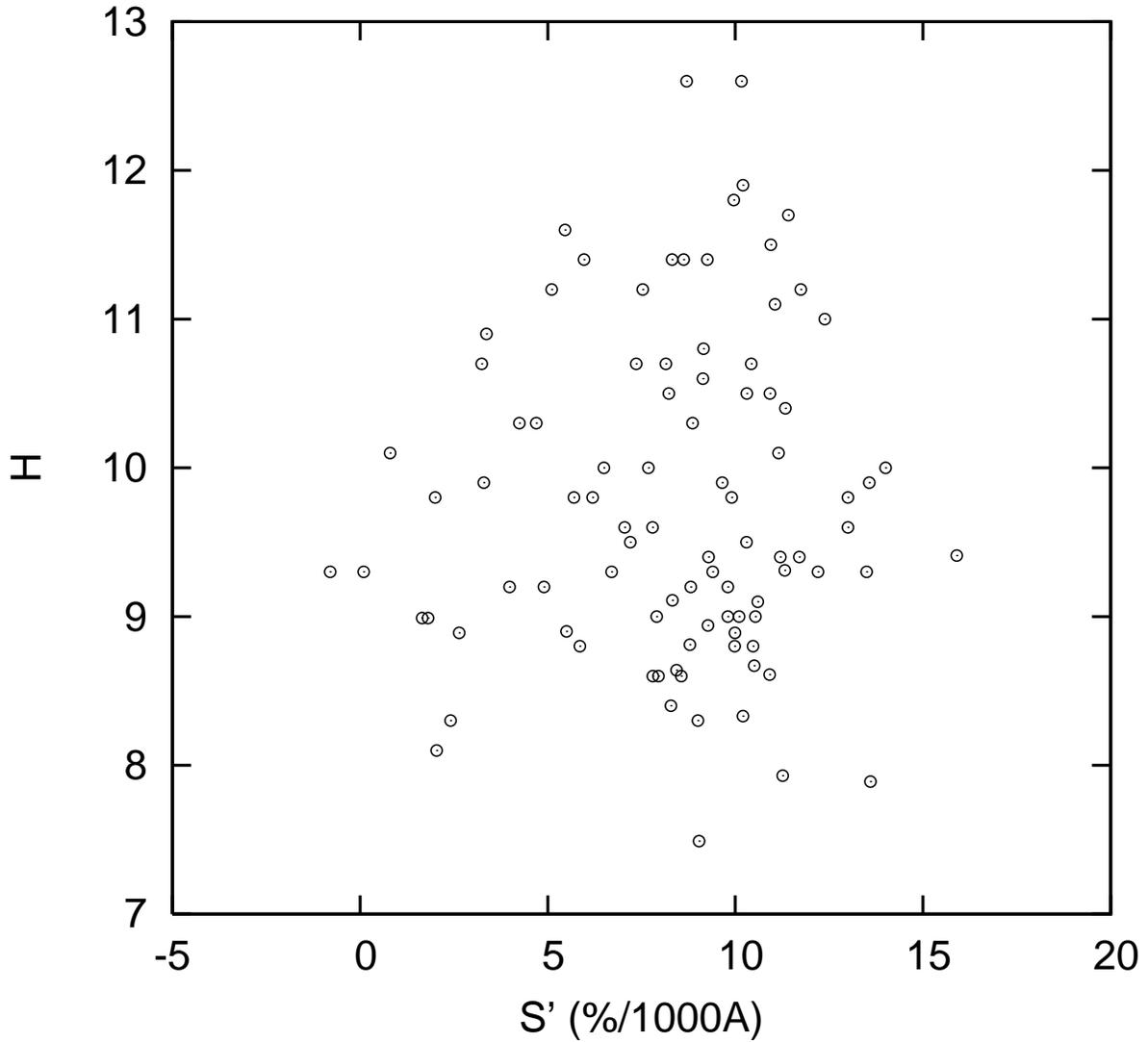}
\caption{Absolute magnitude, H, vs the normalised slope of the visible
spectra, $S'$ for all the available observations.}
\label{fig:HS}
\end{figure}

In figure~\ref{fig:HS} we plot the Absolute magnitude, $H$, against the
normalised slope of the visible spectra, $S'$ for all the observations. It
is apparent the lack of correlation between both quantities. We have
computed the linear correlation coefficient between $H$ and $S'$ as:
$$r_{HS'} = \frac{ \sum_i^N{ (H_i - \overline{H})(S'_i - \overline{S'}) } }{
\sqrt{\sum_i^N{ (H_i - \overline{H}) }}\ \sqrt{ \sum_i^N{(S'_i - \overline{S'})}} },$$
where $N$ is the total number of object with known value of $S'$, which is
$93$ in our case, $\overline{H}$ is the mean value of $H$ and $\overline{S'}$ is  mean
value of $S'$. The correlation  cofficient  between $H$ and $S'$ gives a 
value of $r_{HS'} = 0.06$, indicating that there is no correlation. 

\begin{figure}[bp]
\centerline{\includegraphics[width=\textwidth]{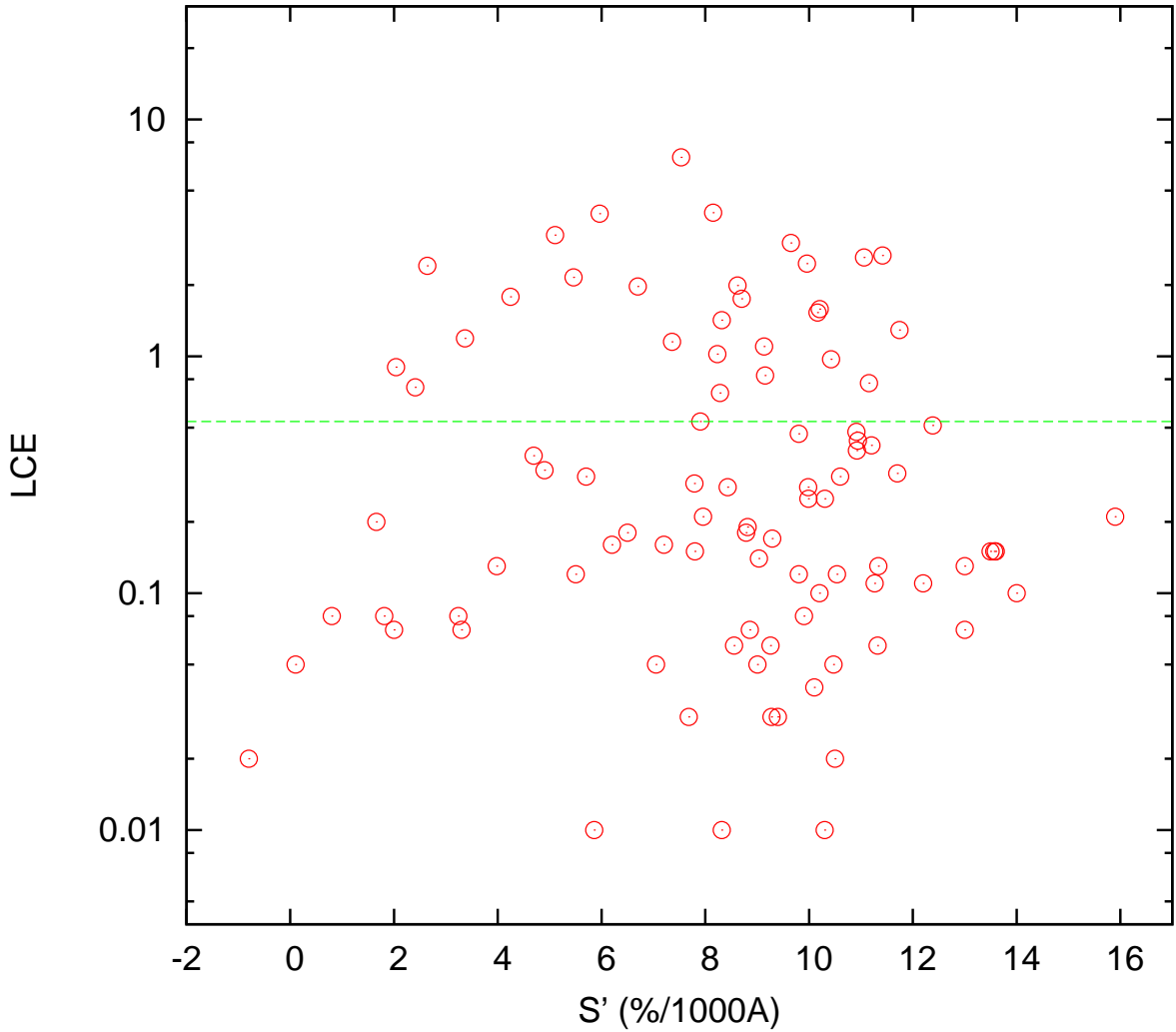}}
\caption{Lyapunov characteristic exponent, $LCE$ (in $1/(10^5 yr)$ units), as a function of the slope
of the visible spectra, $<S'>$ for the Trojan asteroids. Objects above the
horizontal line are considered transitional while the ones below the line
are considered resident. 
}
\label{fig:LCES}
\end{figure}

In Figure~\ref{fig:LCES}, we plot $LCE$ against $S'$. A correlation between
the dynamical stability and the visible color is not apparent, and the 
distributions of slopes of both the transitional and the resident groups
have a similar wide range. As in section~\ref{sec:corrsize}, we divide this
sample into 2 groups, the {\it resident} Trojans have $LCE_{res} \le 0.53\
\times 1/(10^5 yr)$, comprising $65$ objects and the {\it transitional}
ones, $LCE_{trans} > 0.53\ 1/(10^5 yr)$, with $28$ objects. 

\begin{figure}[bp]
\centerline{\includegraphics[width=\textwidth]{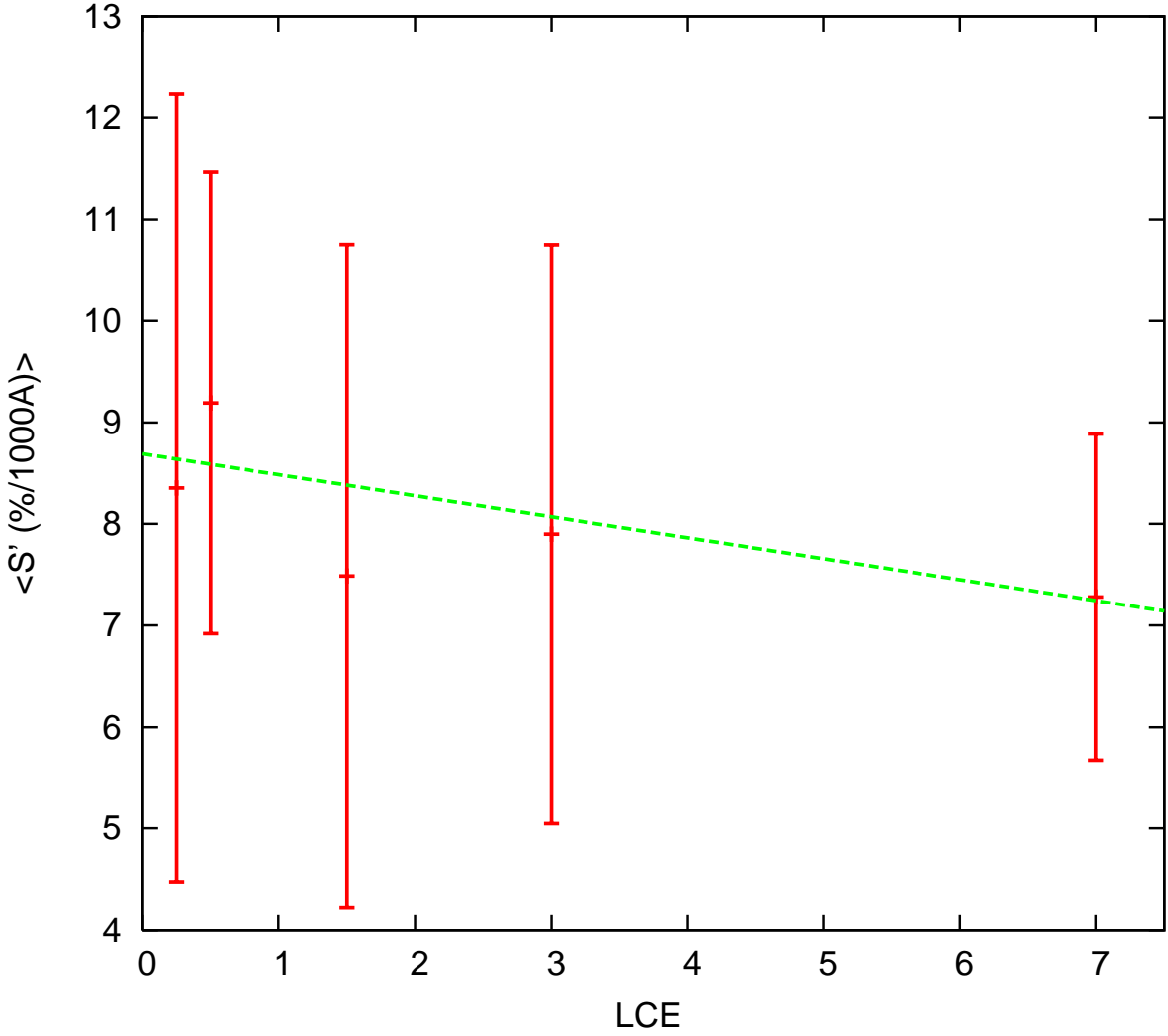}}
\caption{Mean value of the slope of the visible spectra, $<S'>$ as a function
of $LCE$ (in $1/(10^5 yr)$ units). The corresponding dispersions are indicated by the error bars.
The straight line corresponds to a minimum-squares linear fit.  }
\label{fig:LCEHH}
\end{figure}

The probability, $p_{KS}$, that the samples of visible colors,
corresponding to the unstable and the stable groups, are drawn from the same
distribution, given by a Kolmogorov-Smirnov test, is very small:
$p_{KS}=0.17$. This result is probably due to the small size of the sample.
The median value, $<S'>_{med}$, corresponding to the transitional objects is
slightly more neutral than the one corresponding to the residents (see
table~\ref{tab:TT}), although both values are similar within the error
bounds. If we divide the sample according to the value of $LCE$ into more
than $2$ groups, there is a weak tendency for the more stable objects to be
redder (see figure~\ref{fig:LCEHH}). 

\begin{figure}[bp]
\centerline{\includegraphics[width=\textwidth]{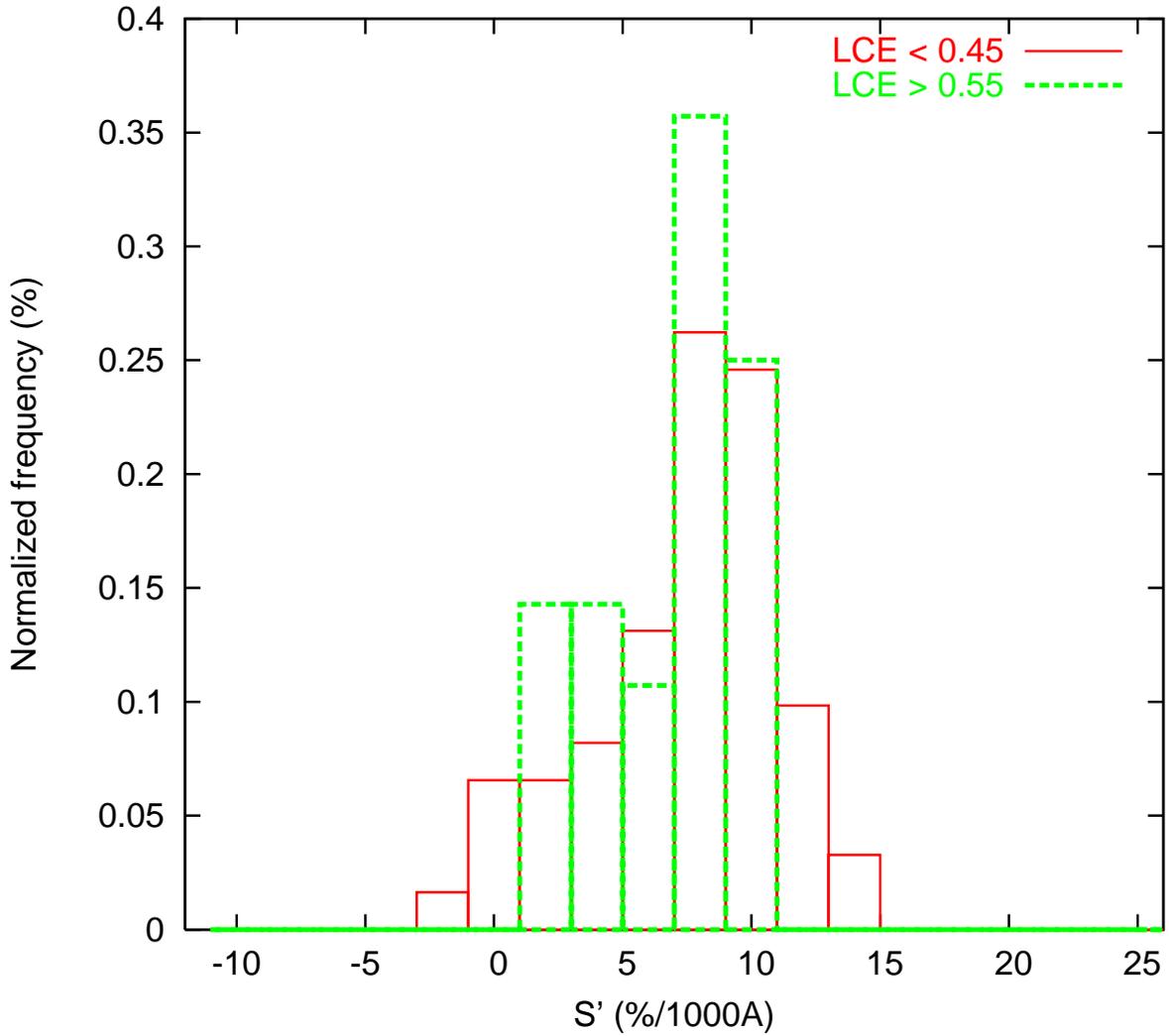}}
\caption{Histograms of $S'$ for both the Resident and the Transitional groups.}
\label{fig:histoS}
\end{figure}

A slight excess of neutral objects exists in the transitional group (see
figure~\ref{fig:histoS}). As we have discussed in
section~\ref{sec:corrsize}, transitional Trojans can be associated with
collisional by-products, which are expected to have younger surfaces. This
result might indicate that Trojan young surfaces are neutral and they tend
to redden with age.

\begin{table}[bp]
\begin{tabular}{|l|c|c|}\hline
Group & $<S'>_{med} (\% / 1000\AA)$ & $\sigma_{S'} (\% / 1000\AA)$ \\ \hline
Resident &  9.18  & 3.22 \\
Transitional & 8.18  & 2.63  \\ \hline
\end{tabular}\\
\caption{Median value of the slope of the visible spectra, $<S'>$ and its
dispersion, $\sigma_{S'}$, for the Resident and the Transitional groups.}
\label{tab:TT}
\end{table}

To study correlations between albedo and color or orbital stability, 
we used the IRAS geometrical albedos (Tedesco \etal 2002) and
those obtained by Fern\'andez \etal (2003), corresponding to the $48$
Trojan asteroids that were in our spectroscopic sample. We did not find any
apparent relationship between those parameters.




We did not find either any statistical differences in the samples of $S'$,
corresponding to the L4 and L5 clouds, composed of $50$ and $43$ objects
respectively (see
Table~\ref{tab:L4L5}). 

\begin{table}[bp]
\begin{tabular}{|l|c|c|c|}\hline
Cloud & $<S'>_{med}$ & $\sigma_{S'}$ & $N$\\ \hline
L4 &   7.82 &  3.54 & 47\\
L5 &  7.58 & 3.60 & 46\\ \hline
\end{tabular}\\
\caption{Median value of the slope of the visible spectra, $<S'>_{med}$ and its
dispersion, $\sigma_{S'}$, and the size of each sample, $N$, for the L4 and L5 clouds.}
\label{tab:L4L5}
\end{table}

\section{Discussion}
\label{sec:disc}

Before performing the observation presented here, we found only $4$ Trojan
asteroids with determined visual spectra in orbits with $LCE>0.53 \times
1/(10^5 yr)$ (see tables~\ref{tab:Spout1}, \ref{tab:Spout2} and
\ref{tab:Spout3}). Therefore, our data has increased by a factor of five the
number of measured visible-spectra of Trojans on unstable orbits. Using this
increased sample, we have studied the distribution of visible-spectroscopic
observations of the Trojan asteroids and search for correlations with
orbital stability, size and albedo.

Some physical characteristics, such as the size distribution, are expected 
to be related with orbital stability, due to energy equipartition after a
physical collision. We find that, while the absolute-magnitude distribution
of the Trojans in stable orbits is bimodal, the one corresponding to the
unstable orbits is unimodal, with a slope similar to the one for the small
stable Trojans. Given the location of the node for the bimodal distribution
of the stable objects and the smallest absolute magnitude of the
transitional ones, we deduce that fragments of collision become more
abundant at $H$ $\gtrsim$ $9.4$. On the other hand, we find no correlation
between size and spectral slope. The resident and the transitional (smaller)
objects, do not differ statistically in their surface properties. Therefore,
we conclude that the collisional process that feeds the unstable orbits and
creates smaller bodies is not altering noticeably the distribution of
surface properties.

\section*{References}


Brunetto, R., Barucci, M. A., Dotto, E. and Strazzulla, G. 2006a. 
Ion Irradiation of Frozen Methanol, Methane, and Benzene: Linking to the
Colors of Centaurs and Trans-Neptunian Objects. {\it The Astrophysical
Journal}, {\bf 644}, 1, 646-650.

Brunetto, R., Romano, F., Blanco, A., Fonti, S., Martino, M., Orofino, V.,
Verrienti, C. 2006b. Space weathering of silicates simulated by nanosecond
pulse UV excimer laser. {\it Icarus}, {\bf 180}, 2, 546-554. y

Bus, S.J. 1999. Compositional structure in the asteroid belt: results of a
spectroscopic survey. {\it PhD Thesis}, MIT, Cambridge.

Bendjoya.,P, Cellino, A, Di Martino, M. and Saba, L. 2004. Spectroscopic
observations of Jupiter Trojans. {\it Icarus}, {\bf 168}, 2, 374-384. 

Chambers, J.E. 1999. A Hybrid Symplectic Integrator that Permits Close
Encounters between Massive Bodies. {\it M.N.R.A.S.}, {\bf 304}, 793-799.

Cruikshank, D. P., Dalle Ore, C. M., Roush, T. L., Geballe,
T. R., Owen, T. C., de Bergh, C., Cash, M. D.;
Hartmann, W. K. 2001. Constraints on the Composition of Trojan
Asteroid 624 Hektor. {\it Icarus}, {\bf 153}, Issue 2, 348-360.

Dell'Oro, A., Marzari, P., Paolicchi F., Dotto, E., Vanzani, V. 1998. Trojan
collision probability: a statistical approach. {\it Astronomy and
Astrophysics}, {\bf 339}, 272-277.

Dotto, E., Fornasier, S., Barucci, M. A., Licandro, J., Boehnhardt, H.,
Hainaut, O., Marzari, F., De Bergh, C., De Luise, F. 2006. The surface
composition of Jupiter Trojans: Visible and Infrared survey of dynamical
Families. {\it Icarus}, {\bf 183}, 2, 420-434.

Dvorack, R., Tsiganis, K. 2000. Why do Trojan ASCS (not) escape?.  
{\it Celestial
Mechanics and Dynamical Astronomy}, {\bf 78}, 125-136.

Emery, J.~P. and Brown, R.~H. 2003. Constraints on the surface composition
of Trojan asteroids from near-infrared (0.8-4.0 Î¼) spectroscopy {\it
Icarus}, {\bf 164}, 1, 104-121.

Emery, J.~P. and Brown, R.~H. 2004. The surface composition of Trojan
asteroids: constraints set by scattering theory. {\it Icarus}, {\bf 170}, 1,
131-152.

Emery J.P., Cruikshank D.P. and Van Cleve J. 2006. Thermal emission
spectroscopy (5.2-38 $\mu$¼) of three Trojan asteroids with the Spitzer
Space Telescope: Detection of fine-grained silicates. {\it Icarus}, {\bf 82},
2, 496-512.

Fern{\' a}ndez, Y.~R. and Sheppard, S.~S. and Jewitt, D.~C. 2003.  The
Albedo Distribution of Jovian Trojan Asteroids. {\it The Astronomical Journal},
{\bf 126}, 1563-1574.

Fornasier, S., Dotto, E., Marzari, F., Barucci, M.~A.,  
Boehnhardt, H. and Hainaut, O. and de Bergh, C., 2004, Visible spectroscopic
and photometric survey of L5 Trojans: investigation of dynamical families,
{\it Icarus}, {\bf 172}, 221-232.

Fornasier, S., Dotto E., Hainaut O., Marzari, F.,   
Boehnhardt, H., De Luise F. and Barucci M.A. 2007, Visible spectroscopic
and photometric survey of Jupiter Trojans: Final result on dynamical families,
{\it Icarus}, {\bf 190}, 622-642.

Gil-Hutton, R. 2002. Color diversity among Kuiper belt objects: The
collisional resurfacing model revisited. {\it Planetary and Space Science}, {\bf
50}, 1, 57-62.

Holsapple K.A. 1993. The scaling of impact processes in planetary sciences.
{\it Annu. Rev. Earth Planet. Sci.}, {\bf 21}, 333-374.

Horner J, Evans N.W. and Bailey M.E. 2005. Simulation of the Population of
Centaurs II. Individual objects. {\it 
M.N.R.A.S.}, {\bf 355}, 2, 321-329.

Jewitt, D.C. 2002, From Kuiper Belt Object to Cometary Nucleus: The Missing
Ultrared Matter. {\it The Astronomical Journal}, {\bf 123}, 2, 
1039-1049.

Jewitt, David C., Luu, Jane X. 1990. CCD spectra of asteroids. II - The
Trojans as spectral analogs of cometary nuclei. {\it Astronomical Journal}, 
{\bf 100}, 933-944.

Jewitt, David C., Trujillo, Chadwick A., Luu, Jane X. 2000.
Population and Size Distribution of Small Jovian Trojan Asteroids.
{\it The Astronomical Journal}, {\bf 120}, 2, 1140-1147.

Karlsson, O. 2004. Transitional and temporary objects in the Jupiter Trojan
area. {\it Astronomy and Astrophysics}, {\bf 413}, 1153-1161.

Landolt, A. U. 1992. 
UBVRI photometric standard stars in the magnitude range 11.5-16.0 around the
celestial equator. 
{\it Astr. J.} {\bf 104}, 1, 340-371, 436-491.

Lazzaro, D., Angeli, C. A., Carvano, J. M., Mothe-Diniz, T,; Duffard, R,;
Florczak, M. 2004. S3OS2: The visible spectroscopic survey of 820
asteroids. {\it Icarus},   {\bf 172}, 179--220.

Levison, H., Shoemaker, E. M., Shoemaker, C. S. 1997. The dispersal of the
Trojan asteroid swarm. {\bf Nature}, {\bf  385},  42-44.

Marzari, F., Farinella, P., Vanzani, V.
Are Trojan collisional families a source for short-period comets?. 1995.
{\it Astronomy and Astrophysics}, {\bf 299}, 267.

Marzari, F. and Scholl, H. 1998. Capture of Trojans by a Growing
Proto-Jupiter {\it Icarus}, {\bf 131}, 1, 41-51.

Milani, A., Nobili, A.M. 1992. An example of stable chaos in the Solar
System {\it Nature}. {\bf 357}, 6379, 569-571. 

Milani, A. 1993.  The Trojan asteroid belt: Proper elements, stability,
chaos and families {\it Celestial Mechanics and Dynamical Astronomy}, {\bf
57}, 1-2, 59-94.

Milani, A., Nobili, A.M., Knezevic, Z. 1997.  Stable Chaos in the Asteroid Belt. 
{\it Icarus}, {\bf 125}, 1, 13-31. 

Morbidelli, A., Levison, H. F., Tsiganis, K. and Gomes, R. 2005.
Chaotic capture of Jupiter's Trojan asteroids in the early Solar System
{\it Nature}, {\bf 435}, 7041, 462-465.

Moroz, L., Baratta, G., Strazzulla, G., Starukhina, L.,
Dotto, E., Barucci, M.A., Arnold, G., Distefano,
E. 2004. {\it Icarus}, {\bf 170}, 1, 214-228.

Nesvorny, D. and Dones, L. 2002. How Long-Lived Are the Hypothetical Trojan
Populations of Saturn, Uranus, and Neptune? {\it Icarus}, {\bf 160}, 2,
271-288. 

Petit, J.M. and Farinella, P. 1993.  Modelling the outcomes of
high-velocity impacts between small solar system bodies. {\it Celestial
Mechanics and Dynamical Astronomy}, {\bf 57}, 1-2, 1-28.

Pilat-Lohinger, E.,  Dvorak, R. 1999. 
Trojans in Stable Chaotic Motion. {\it Celestial
Mechanics and Dynamical Astronomy}, {\bf 73}, 117-126.

Rabe, E. 1972. Orbital Characteristics of Comets Passing Through the 1:1
Commensurability with Jupiter. The Motion, Evolution of Orbits, and Origin
of Comets, Proceedings from IAU Symposium no. 45, held in Leningrad,
U.S.S.R., August 4-11, 1970. Edited by Gleb Aleksandrovich Chebotarev, E. I.
Kazimirchak-Polonskaia, and B. G. Marsden. International Astronomical Union.
Symposium no. 45, Dordrecht, Reidel, p.55.

Strazzulla, G., Dotto, E., Binzel, R., Brunetto, R., Barucci, M. A., Blanco,
A., Orofino, V. 2005. Spectral alteration of the Meteorite Epinal (H5)
induced by heavy ion irradiation: a simulation of space weathering effects
on near-Earth asteroids. {\it Icarus}, {\bf 174}, 1, 31-35. 


Tedesco, E.F., Noah, P.V., Noah, M., Price and Stephan D. 2002. 
The Supplemental IRAS Minor Planet Survey.
{\it The Astronomical Journal}, {\bf 123}, 2, 1056-1085.

Tsiganis, K., Dvorak, R., Pilat-Lohinger, E. 2000.
Thersites: a jumping' Trojan?
{\it Astronomy and Astrophysics}, {\bf 354}, 1091-1100.

Tsiganis, K., Varvoglis, H., Dvorak, R. 2005.  Chaotic Diffusion And
Effective Stability of Jupiter Trojans {\it Celestial Mechanics and
Dynamical Astronomy}, {\bf 92}, 1-3. 71-87

Yang, B, Jewitt, D. 2007. Spectroscopic Search for Water Ice on Jovian Trojan
Asteroids. {\it The Astronomical Journal}, {\bf 134}, 1, 223-228.

Yoder, C. F. 1979. Notes on the origin of the Trojan asteroids. {\it Icarus},
{\bf  40}, 341-344.

\end{document}